\documentclass[sn-mathphys-num]{sn-jnl}

\usepackage{graphicx}%
\usepackage{multirow}%
\usepackage{amsmath,amssymb,amsfonts}%
\usepackage{amsthm}%
\usepackage{mathrsfs}%
\usepackage[title]{appendix}%
\usepackage{xcolor}%
\usepackage{textcomp}%
\usepackage{manyfoot}%
\usepackage{booktabs}%
\usepackage{algorithm}%
\usepackage{algorithmicx}%
\usepackage{algpseudocode}%
\usepackage{listings}%

\theoremstyle{thmstyleone}%
\theoremstyle{thmstyletwo}%

\theoremstyle{thmstylethree}%

\begin{document}
\title{Zak Phase Induced Topological Nonreciprocity}


\author[1]{\fnm{Xiao} \sur{Liu}}
\equalcont{These authors contributed equally to this work.}

\author[1,2]{\fnm{Jiefei} \sur{Wang}}
\equalcont{These authors contributed equally to this work.}

\author[1]{\fnm{Ruosong} \sur{Mao}}

\author[2]{\fnm{Huizhu} \sur{Hu}}

\author[1,2,3]{\fnm{Shi-Yao} \sur{Zhu}}

\author*[1]{\fnm{Xingqi} \sur{Xu}}\email{xuxingqi@zju.edu.cn}

\author*[2]{\fnm{Han} \sur{Cai}}\email{hancai@zju.edu.cn}

\author*[1,2,3]{\fnm{Da-Wei} \sur{Wang}}\email{dwwang@zju.edu.cn}

\affil[1]{\orgdiv{Zhejiang Key Laboratory of Micro-Nano Quantum Chips and Quantum Control}, \orgname{School of Physics, and State Key Laboratory for Extreme Photonics and Instrumentation, Zhejiang University}, \orgaddress{\city{Hangzhou}, \postcode{310027}, \state{Zhejiang Province}, \country{China}}}

\affil[2]{\orgdiv{College of Optical Science and Engineering}, \orgname{Zhejiang University}, \orgaddress{\city{Hangzhou}, \postcode{310027}, \state{Zhejiang Province}, \country{China}}}

\affil[3]{\orgdiv{Hefei National Laboratory}, \orgaddress{\city{Hefei}, \postcode{230088}, \state{Anhui Province}, \country{China}}}

\abstract{Topological physics provides novel insights for designing functional photonic devices, such as magnetic-free optical diodes, which are important in optical engineering and quantum information processing. Past efforts mostly focus on the topological edge modes in two-dimensional (2D) photonic Chern lattices, which, however, require delicate fabrication and temporal modulation. In particular, the 1D nonreciprocal edge mode needs to be embedded in a 2D lattice, contradicting with the compactness of integrated photonics. To address these challenges, we investigate the  optical nonreciprocity of the 1D Su-Schrieffer-Heeger (SSH) superradiance lattices in room-temperature atoms. The probe fields propagating in two opposite directions perceive two different SSH topological phases, which have different absorption spectra due to the interplay between the Zak phase and the thermal motion of atoms, resulting in optical nonreciprocity. Our findings reveal the relationship between 1D topological matter and optical nonreciprocity, simplifying the design of topologically resilient nonreciprocal devices.}

\maketitle

\newpage 
\section*{Introduction}

Optical nonreciprocity (ONR) \cite{Jalas2013,Sounas2017,Zhang2023,Khanikaev2010} has diverse applications in optical engineering \cite{Estep2014,Kang2011,Kim2015,Chang2014,King2024,Cotrufo2024} and quantum information processing \cite{Yu2009,Lodahl2017,Kimble2008}. In the linear regime, ONR needs the  time reversal symmetry breaking, which is usually achieved by the Faraday effect. However, the incorporation of magnetic fields can compromise the miniaturization of the device and its integration with the electronic devices. Topological physics \cite{Hasan2010,Qi2011,Ozawa2019} provides a nonmagnetic way of breaking the time-reversal symmetry by using chiral edge states in two-dimensional (2D) photonic materials \cite{Fang2012,Fang2012PRL,Tzuang2014}. Similar to the quantum Hall effect, these edge states are protected by the non-zero Chern number of the 2D Bloch bands, making them robust against local defects. Utilizing the analogy between the Schr\"{o}dinger equation and Maxwell's equations \cite{Haldane2008,Raghu2008}, such chiral edge modes have been realized in arrays of microwave resonators \cite{Wang2009} and photonic waveguides \cite{Rechtsman2013}. 

These 2D topological lattices require an extra dimension of resources and sophisticated temporal modulation \cite{Sounas2017} to break the time-reversal symmetry. The miniaturization of the 2D bulk of these lattices is limited by the edge to edge scattering \cite{Mancini2015,Stuhl2015,Cai2019,Chalopin2020}. To address this challenge, we resort to the 1D topological models, such as the Su-Schrieffer-Heeger (SSH) model, which are characterized by the geometric Zak phase \cite{Zak1989,Xiao2010,Mao2022,Atala2013,Meier2018,Li2023}, to simplify the design  and improve the utilization rate of the topological materials. However, the SSH model involves with the inversion symmetry breaking, rather than the time-reversal symmetry breaking required by the ONR. In addition, we cannot directly use the edges states in the SSH models since they are localized and nonpropagating. It underscores the need of new ingredients to harness the 1D topological phases to obtain ONR.

To use the two topological phases of the SSH model to achieve ONR, we need to transform the inversion symmetry breaking to time-reversal symmetry breaking, and the two topological phases need to have different optical absorption. The former can be achieved by using a momentum-space lattice, where the lattice sites are composed by states possessing different momenta. In such lattices, the inversion of momentum is equivalent to the reversion of time, such that the inversion symmetry breaking means the time-reversal symmetry breaking. The latter can be achieved by using the modern theory of polarization, which relates the topological Zak phases to the optical response of materials. These two conditions can be simultaneously satisfied in the superradiance lattices (SLs) \cite{Wang2015}, which are momentum-space lattices of timed Dicke states \cite{Scully2006,He2020} and the topological Zak phases have been demonstrated in their absorption spectra \cite{Mao2022}. 

 Here we report the experimental realization of topological ONR in an SSH SLs  of thermal atoms. Such lattices are constructed in an electromagnetically induced transparency (EIT) configuration where the coupling field is a partial standing wave with two unequal plane wave components.
Compared to previous approaches of ONR in atomic platforms, such as those due to moving photonic crystals \cite{Wang2013,Horsley2013} and atomic collisions \cite{Liang2020}, our method is protected by the two topological phases of the underlying SSH model \cite{Su1979}. These phases are determined solely by the relative field strength of the two plane wave coupling fields. Particularly, in the limit where one of the coupling fields is completely turned off, our system reduces to the previous experiments involving with the so-called susceptibility-momentum locking \cite{Zhang2018}.

\section*{Results}

\begin{figure*}[htbp]
    \centering
    \includegraphics[width=0.8\textwidth]{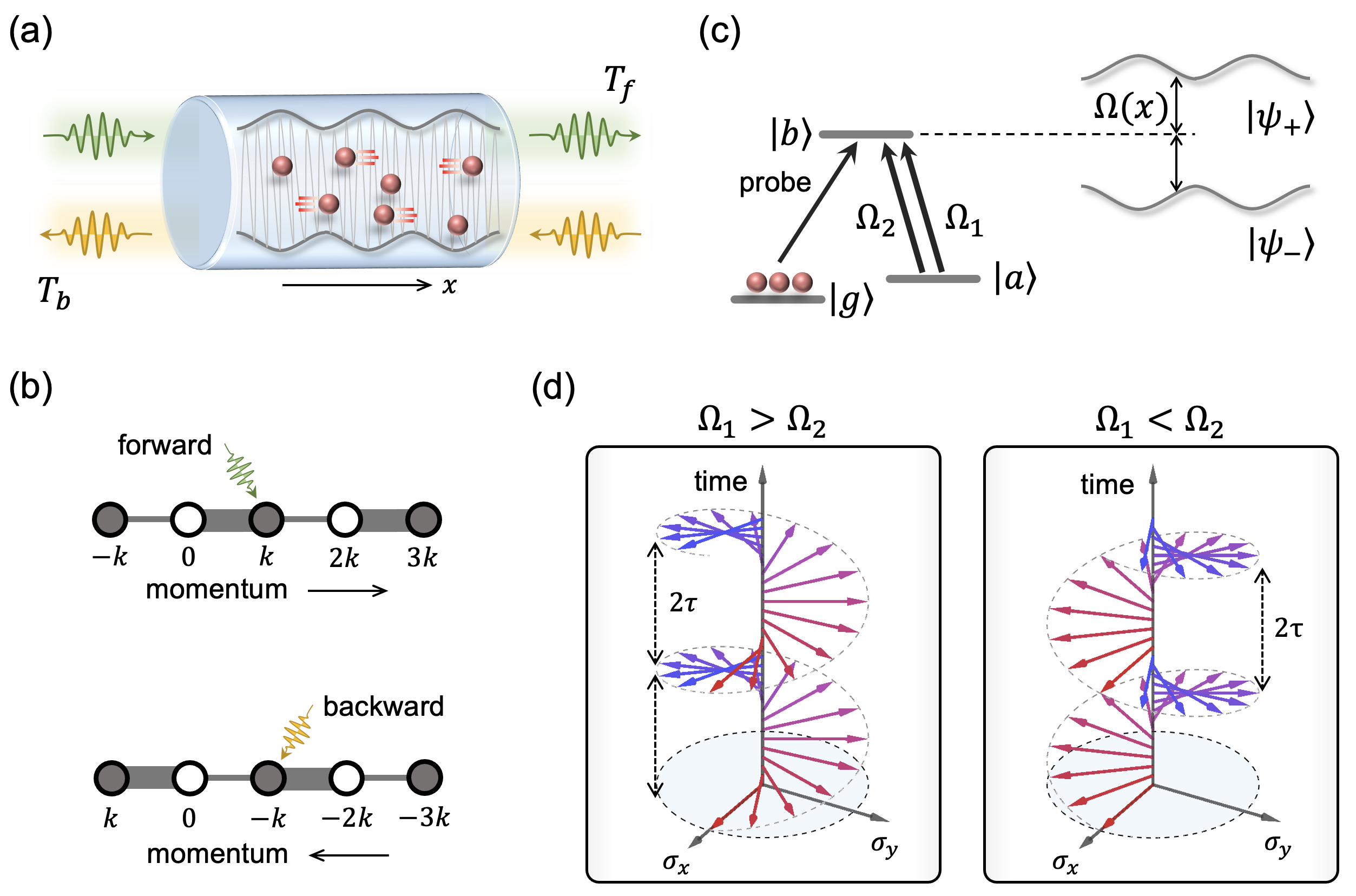}
    \caption{
        \textbf{Schematics of Zak phase induced optical nonreciprocity.} (a) The experimental setup. Thermal atoms in a vapor cell are coupled by a partial standing wave. Two probe fields probe atoms in EIT from two opposite directions. The ONR is characterized by the transmission contrast in two opposite directions. (b) Two different topological phases of the SSH SLs probed by the forward and backward probe fields. Each site is a timed Dicke state (filled circles for $|b\rangle$ and empty circles for $|a\rangle$) with discrete momenta. The horizontal arrows indicate the direction of increasing the momentum. The wave like arrows indicate the sites being probed. By inverting the momentum, the two coupling strengths interchange with each other. (c) The atomic level configuration and the partial standing-wave coupling field induced periodic level splitting. (d) Rotation of the Bloch vector of the eigenstates on the equator of the Bloch sphere when the atom is moving through the standing wave. In this process, the atoms acquire a Zak phase $\theta$ and thus have an effective energy shift $\Delta_Z$. Here $\sigma_i$ ($i=x,y$) stands for the Pauli matrices of the atom in the basis of $|a\rangle$ and $|b\rangle$. The color of the Bloch vector is used to guide the eyes.}
        \label{fig1}
\end{figure*}


 We implement our experiment in a vapor cell containing natural abundance rubidium atoms, as schematically shown in Fig.~\ref{fig1}(a).  Three hyperfine levels of the $^{87}$Rb D1 line are coupled in a $\Lambda$-type EIT configuration. The probe field couples the atomic transition between the ground state $|g \rangle\equiv|5{^2S_{1/2}},F=1 \rangle$ and the excited state $|b \rangle\equiv|5{^2P_{1/2}},F=2 \rangle$. The transmissions of the probe fields propagating along the forward ($+x$) and backward ($-x$) directions are $T_{f}$ and $T_{b}$, which are used to quantify the ONR with the contrast $\eta=(T_{f}-T_b)/(T_f+T_b)$. Two counter-propagating coupling fields resonantly drive the atomic transition between the metastable state $|a \rangle\equiv|5{^2S_{1/2}},F=2 \rangle$ and  $|b \rangle$, leading to the effective Hamiltonian $H_c(x)= \Omega(x)|a\rangle\langle b|+H.c.$,
where $\Omega(x)\equiv |\Omega(x)|e^{i\phi}=\Omega_1 e^{ikx}+\Omega_2 e^{-ikx}$, $\Omega_{1}$ ($\Omega_{2}$) is the Rabi frequency of the $+x$ ($-x$) plane wave component of the coupling fields and  $k=2\pi/\lambda$ is the coupling field wavevector.
It is easy to obtain 
$|\Omega(x)|=[{(\Omega_1+\Omega_2)^2\cos^2{kx}+(\Omega_1-\Omega_2)^2\sin^2{kx}}]^{1/2}$ and $\tan\phi=\frac{\Omega_1-\Omega_2}{\Omega_1+\Omega_2}\tan kx$. Therefore, $\phi$ increases with $x$ when $\Omega_1>\Omega_2$ but decreases otherwise, which will be shown to determine the sign of the Zak phase. 

In momentum space, the timed Dicke states created by the probe fields are coupled by the two plane wave coupling fields with two different coupling strengths $\Omega_1$ and $\Omega_2$ to form an SSH SL \cite{Wang2015,Chen2018,Mi2021} (see Methods). The topological phases of the SSH SLs are determined by the relative strengths of $\Omega_1$ and $\Omega_2$. The time-reversal symmetry is broken because the backward probe field sees a lattice the same as the forward probe field but only by interchanging $\Omega_1$ and $\Omega_2$, as schematically shown in Fig.~\ref{fig1}(b). Therefore, the two probe fields see two different topological phases of the SSH model. However, we still need the two topological phases to have different absorption to achieve ONR. Interestingly, the thermal motion of the atoms projects the topological Zak phases in the absorption spectra of the SLs, such that the two topological phases indeed have different absorption for a variety of probe frequencies. We elaborate this point in the following.

The interplay between the Zak phase and the atomic motion can be better understood in real space, i.e., the eigenspace of the SLs. Without atomic motion, the eigenstates are $|\psi_\pm(x)\rangle =\left(|a\rangle\pm e^{-i\phi(x)}|b\rangle\right)/\sqrt{2}$, where the subscripts ``$+$'' and ``$-$'' represent the upper and lower energy bands, as shown in Fig.~\ref{fig1}(c). However, the atomic motion along the $\pm x$ direction introduces a Bloch oscillation in the SLs. This is similar to the dynamics of a crystal electron in a static electric field. The atoms periodically go through the real space Brillouin zone determined by the standing wave pattern. As a result, the energy spectra transit from energy bands to Wannier-Stark ladders (WSLs)  \cite{Zak1989,Xiao2010,Mao2022,Sundaram1999,King1993,Resta1994} with separations determined by the Bloch oscillation frequency ($2kv$) and locations determined by the Zak phases.

 Specifically, for the atoms moving with a velocity $v$, it takes the time $\tau=\lambda/2v$ to travel through the period of a partial standing wave, $\lambda/2$. In such a process the eigenstate adiabatically evolves along the equator of the Bloch sphere (as shown in Fig.~\ref{fig1}(d)) and accumulates a geometric Zak phase 
\begin{equation}
\begin{split}
    \theta &= i\int_0^{\lambda/2} dx  \langle \psi_{\pm}|\frac{\partial }{\partial x}|\psi_{\pm}\rangle= \frac{1}{2}\int_0^{\lambda/2} dx  \frac{\partial \phi}{\partial x} =\frac{\xi\pi}{2},
\end{split}
\label{theta}
\end{equation}
where $\xi\equiv\text{sgn}(\Omega_1-\Omega_2)$ determines the direction of the rotation. 
After a time $t$, the total accumulated phase is $\theta t/\tau$ and the states are attached with a  phase factor 
\begin{equation}
    \exp\left[{i\theta t/\tau}\right] = \exp\left[{i \xi kv t/2}\right].
\end{equation}
Therefore, the frequencies of the states $|\psi_\pm\rangle$ are shifted by $\Delta_Z=-\xi kv/2$ due to the interplay between the Zak phase and the motion of atoms. 

The Doppler shifts of the forward and backward probe fields are $\Delta_D^f=-kv$ and $\Delta_D^b=kv$. If we only have these Doppler shifts, after we integrate over the symmetric Maxwell velocity distribution, the total absorption shall be the same for the two probe fields. However, after we consider $\Delta_Z$, the total frequency shifts of the probe fields relative to the atomic transition are $\Delta^f=\Delta_D^f-\Delta_Z=(\xi/2-1)kv$ and $\Delta^b=\Delta_D^b-\Delta_Z=(\xi/2+1)kv$, which satisfy the relation $\Delta^b=\Delta^f(\xi\rightarrow -\xi,v\rightarrow -v)$ from the symmetry of the system. Explicitly, for $\xi=1$, we have $\Delta^f=-kv/2$ and $\Delta^b=3kv/2$, while for $\xi=-1$, we have $\Delta^f=-3kv/2$ and $\Delta^b=kv/2$. 

\begin{figure}[t]
    \centering
    \includegraphics[width=0.7\textwidth]{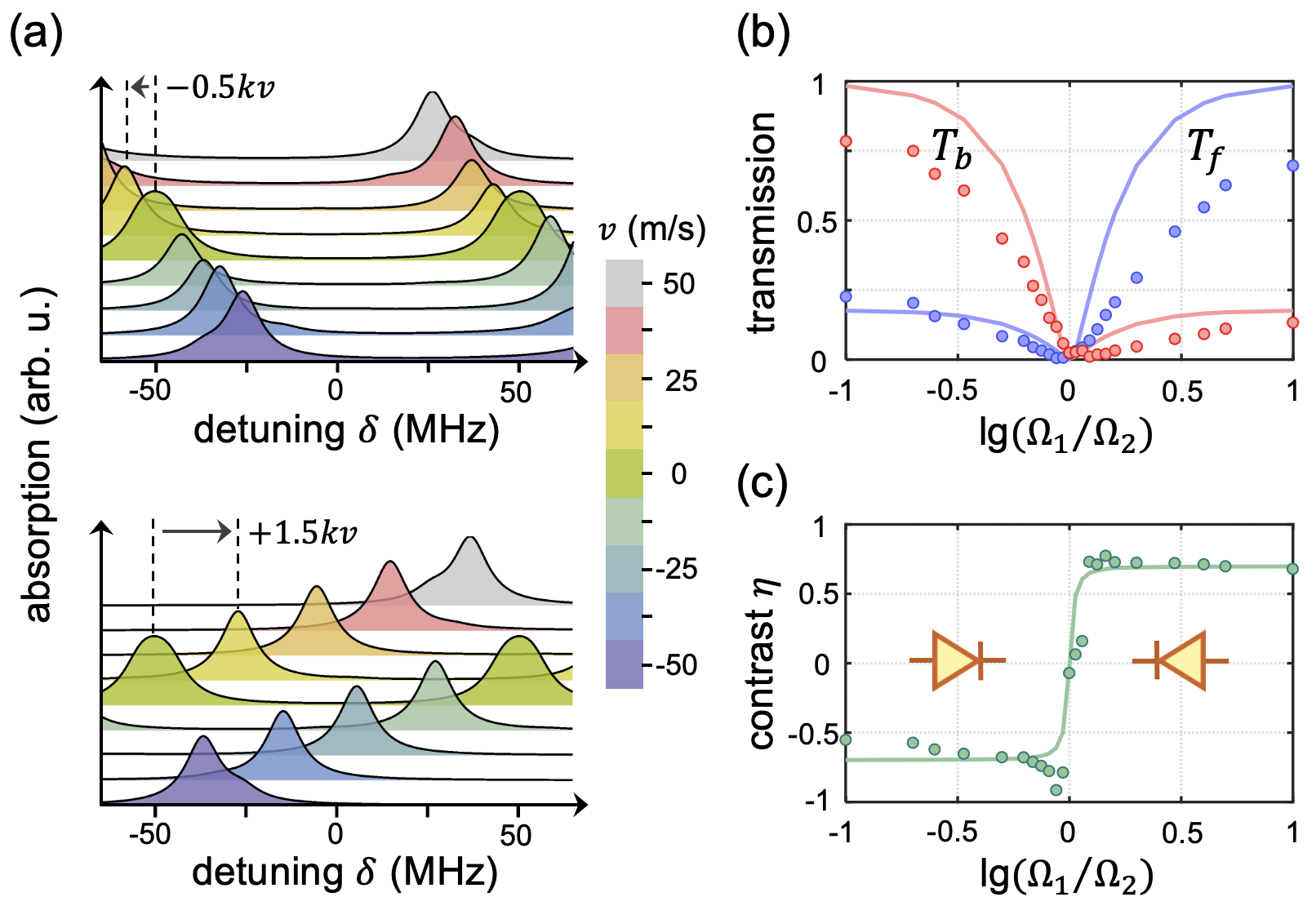}
    \caption{
        \textbf{Optical nonreciprocity in the Su-Schrieffer-Heeger superradiance lattices.} (a) The numerical simulation of atomic velocity-dependent absorption of the forward (upper panel) and the backward (lower panel) probe field. The parameters are $\Omega_1=50$ MHz, $\Omega_2=5$ MHz.
        (b) The probe transmission (upper panel) and contrast (lower panel) as a function of the relative control field strength $\Omega_1/\Omega_2$. The experimental parameter $\Omega_1=136$ MHz when $\Omega_1>\Omega_2$ and $\Omega_2=136$ MHz when $\Omega_1<\Omega_2$. `lg' stands for common logarithm. Temperature of the cell is $70^\circ$C. The lines are numerical results and the dots are experimental data.}
        \label{fig2}
\end{figure}

Since these shifts are still proportional to $v$, we can understand $\Delta^f$ and $\Delta^b$ as the effective Doppler shifts modified by the Zak phase.  The two probe fields see atoms with different frequency broadening, one suppressed and the other enhanced, as shown by the two different slopes in Fig.~\ref{fig2}(a).  Here the detuning $\delta$ is defined as the  atomic transition frequency between $|g\rangle$ and $|b\rangle$  minus the probe field frequency. Therefore, in general, the absorption spectra are different for the two probe fields after we integrate the contributions from all atoms in the Maxwell velocity distribution. As a result, we obtain ONR with transmission direction being determined by the topological phases of the SSH models. 


We take the zero probe detuning as an example to show the relation between the topological phases and the ONR (see Fig.~\ref{fig2}(b)). We observe that $T_f>T_b$ for $\Omega_1>\Omega_2$ and $T_f<T_b$ for $\Omega_1<\Omega_2$. To quantify the ONR, we plot the contrast $\eta$ as a function of $\lg(\Omega_1/\Omega_2)$ in Fig.~\ref{fig2}(c). There is a sharp transition of $\eta$ from negative to positive at $\Omega_1=\Omega_2$. Such a transition can be used to characterize the topological phase transition of the SSH SLs.

\begin{figure*}[ht!]
    \centering
    \includegraphics[width=1\textwidth]{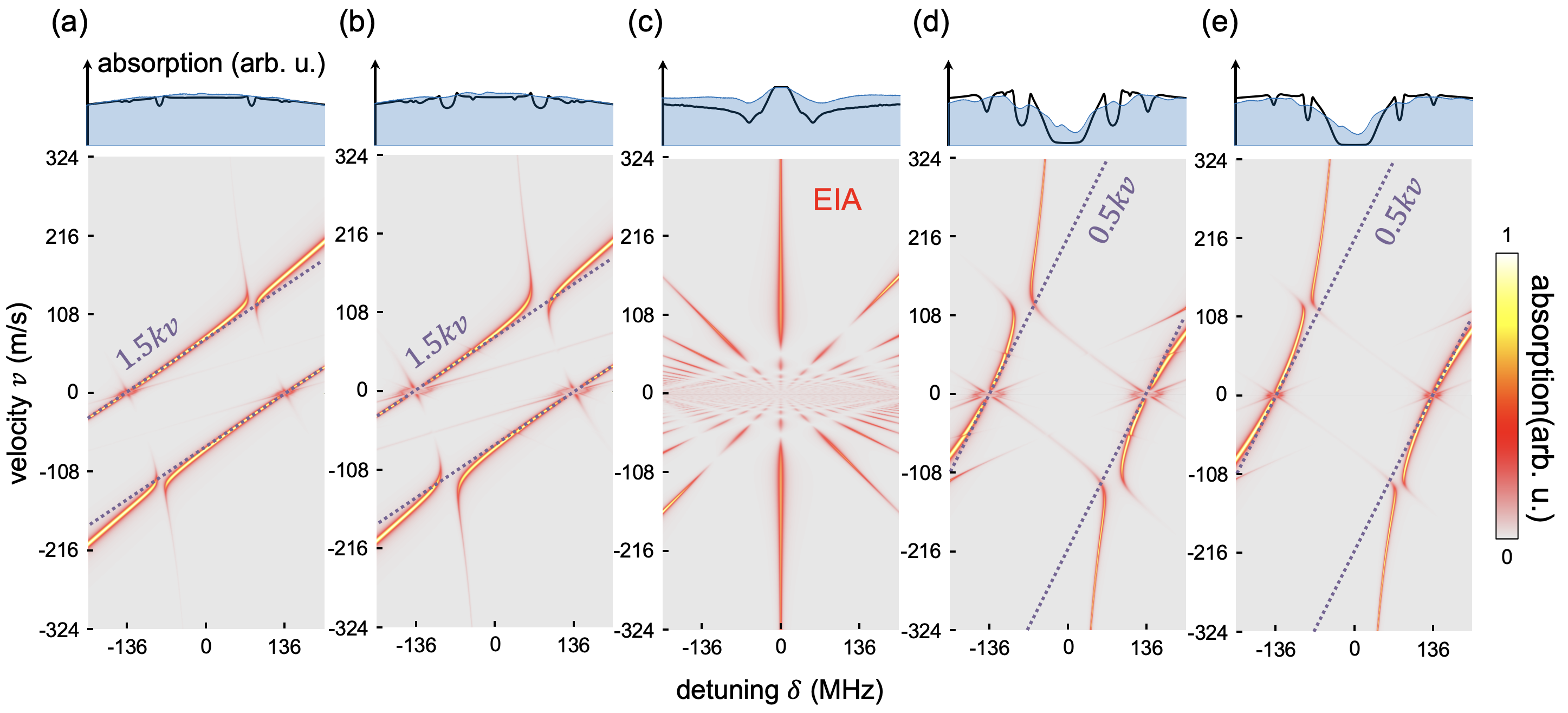}
    \caption{
        \textbf{Absorption spectra of Su-Schrieffer-Heeger superradiance lattices.} The upper panels show the total absorption spectra for the backward probe field and the lower panels show the velocity-dependent ones. In the upper panels, the shaded areas indicate the experimental data and the black lines are the numerical simulation. The dotted lines show the asymptotic eigenenergies for small velocities.
        The Rabi frequencies of the coupling fields are (a) $\Omega_1=136$ MHz, $\Omega_2=14$ MHz; (b) $\Omega_1=136$ MHz, $\Omega_2=27$ MHz; (c) $\Omega_1=136$ MHz, $\Omega_2=136$ MHz; (d) $\Omega_1=27$ MHz, $\Omega_2=136$ MHz; (e) $\Omega_1=14$ MHz, $\Omega_2=136$ MHz. At the topological phase transition point ($\Omega_1=\Omega_2$), the spectrum shows enhanced absorption near zero detuning.}
        \label{fig4}
\end{figure*}

\begin{figure}[t]
    \centering
    \includegraphics[width=0.6\textwidth]{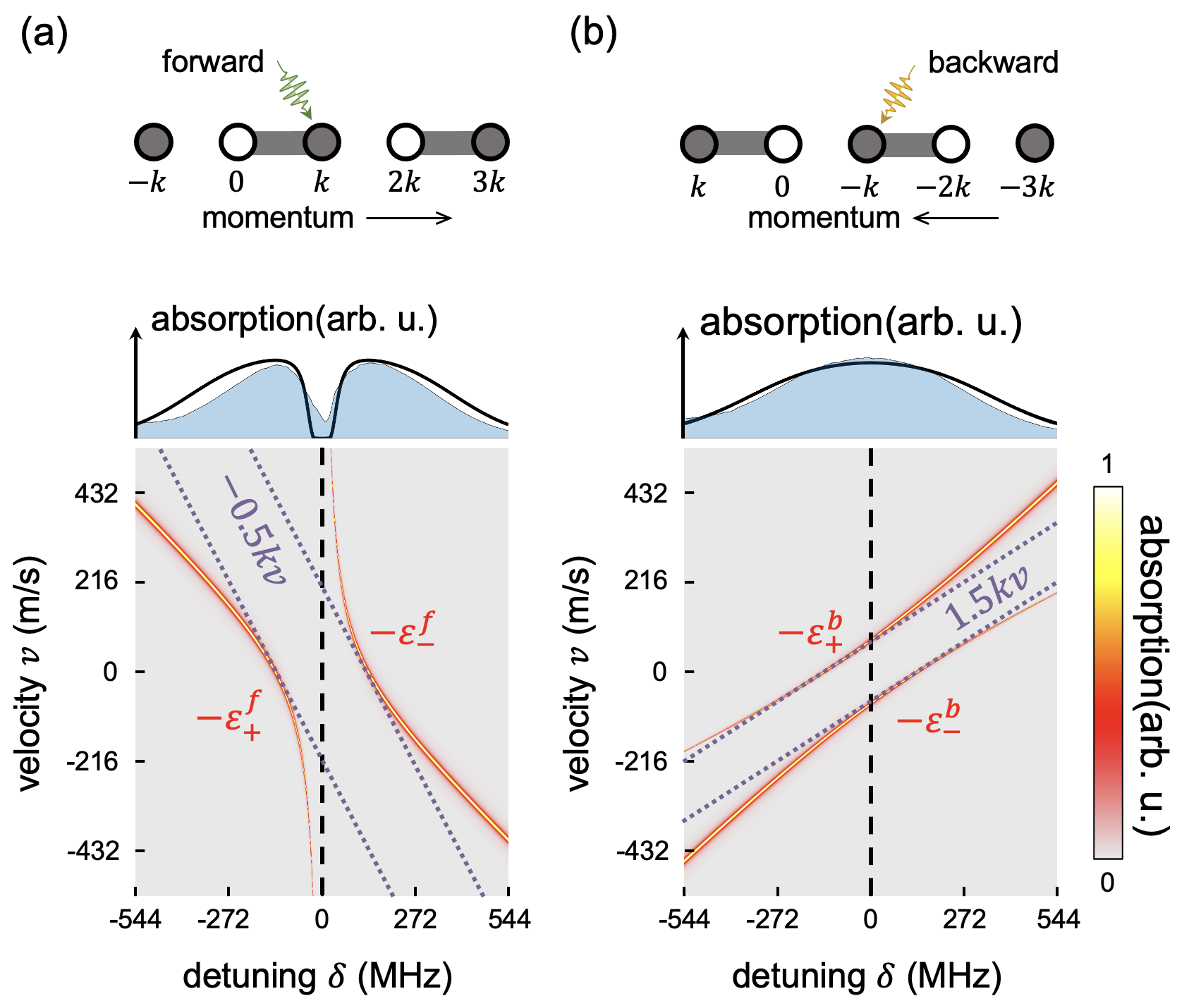}
    \caption{
        \textbf{Optical nonreciprocity in the dimerization limit of the Su-Schrieffer-Heeger superradiance lattices.} (a) Absorption spectra for the forward probe field. (b) Absorption spectra for the backward probe field. The upper panels show the total absorption spectra and the lower panels show the velocity-dependent ones. The absorption follows the eigenenergies with a negative sign due to the definition of $\delta$, i.e., the transition frequency minus the probe frequency. The experimental parameters are $\Omega_1=136$ MHz, $\Omega_2=0$. We observe a transparency window for the forward probe field at the zero probe detuning, in contrast to an absorption peak for the backward probe field.}
        \label{fig3}
\end{figure}

The absorption spectra are characterized not only by the frequency shifts of the energy band centers, $\Delta^f$ and $\Delta^b$, but also by the WSLs, because the atoms periodically pass through the real space Brillouin zone \cite{Mao2022}. For these two-band SLs, each band has a series of WSLs, which have anti-crossing when two ladders from different bands have the same energy for a certain velocity, as shown in Fig.~\ref{fig4}. Despite all these complications, the nonreciprocal absorption is robust for various field strengths, which can be seen by comparing the absorption spectra for two reciprocal values of $\Omega_1/\Omega_2$ (i.e., Fig.~\ref{fig4} (a) and (e); (b) and (d)). For $\Omega_1=\Omega_2$, the absorption at the zero detuning has a strong absorption peak because one of the WSLs for all velocities overlap at this frequency, which can explain the electromagnetically induced absorption \cite{Lezama1999}. The ONR disappears at this topological phase transition point.

At the end, we show that the ONR due to the difference between co-propagating and counter-propagating EIT in thermal atoms \cite{Zhang2018} is a special case of topological ONR in the dimerization limit of the SSH SLs. Here we set $\Omega_2=0$ such that the lattice is dimerized, as shown in Fig.~\ref{fig3}. 
The shifted eigenenergies for the forward probe field is  $\varepsilon^{f}_{\pm}=kv/2\pm\sqrt{(kv)^2/4+\Omega_1^2}$ (see Fig.~\ref{fig3}(a) and Methods). It is interesting to notice that for small and large velocities, the slopes of the eigenenergy as a function of the velocity are fractional and integer, respectively. The fraction comes from the Zak phases. 

For a small velocity such that $|kv|\ll\Omega_1$, we obtain $\varepsilon^f_{\pm}\approx kv/2\pm\Omega_1$ (dotted lines) following the WSL with a perturbative $v$ \cite{Zak1989,Xiao2010,Mao2022,Sundaram1999,King1993,Resta1994}. The slope $\partial \varepsilon^f_{\pm}/\partial (kv)=1/2$ (see Methods), containing the contributions from both the Doppler shift and the Zak phase. For large velocities such that $kv\gg \Omega_1$, we can neglect $\Omega_1$ and $\varepsilon^f_{+}\approx kv$ and $\varepsilon^f_{-}\approx 0$ such that $\partial \varepsilon^f_{+}/\partial (kv)\approx 1$ and $\partial \varepsilon^f_{-}/\partial (kv)\approx 0$. These integer slopes indicate that only the the contribution of the Doppler shift is left, because the Zak phase exists only when $v$ is perturbative. As a result, at the zero detuning (the dashed line), there is no eigenenergy for all velocities. After summing the contribution of all the atoms, such a feature leads to a transparency window at the zero probe detuning. 

For the backward probe field, the eigenenergies $\varepsilon^b_{\pm}=-3kv/2\pm\sqrt{(kv)^2/4+\Omega_1^2}$ can be zero at $kv=\pm\Omega_1/\sqrt{2}$ such that there is absorption at the zero probe detuning, as shown  in Fig.~\ref{fig3}(b). Therefore, the experiment of ONR due to the susceptibility-momentum locking \cite{Zhang2018} can be explained as the dimerization limit of the topological ONR.

\section*{Conclusion}

In summary, we have demonstrated the ONR induced by the 1D topological Zak phase, which is fundamentally different from previous studies based on the chiral edge modes of 2D topological materials.  Compared to previous studies on ONR with moving atoms \cite{Horsley2013} or standing wave \cite{Wang2013}, which break the time-reversal symmetry, we used a static partial standing wave coupling field which breaks the inversion symmetry in a momentum space SL, which is equivalent to the time-reversal symmetry breaking. In momentum space, the two probe fields propagating in opposite directions see topologically distinct SSH SLs.  The 1D topological invariant, i.e., the Zak phase, induces effectively enhanced or suppressed Doppler shifts for the two probe fields, leading to the topologically protected ONR. 

Our results can also be extended to lattices with nontopological Zak phases, since the scheme is based on the inversion symmetry breaking in momentum space lattices. As long as the two probe fields see different Zak phases, we have ONR. We can break the chiral symmetry of the SSH model by detuning the coupling field frequency from the atomic transition \cite{Mao2022}, which introduces sublattice potential difference to the SSH SLs. For example, with $\theta = 0.1\pi$, the modified Doppler shift are $\Delta^f = -0.9kv$ and $\Delta^b = 1.1kv$, respectively, from which we notice that the nonreciprocity still exists but weaker than the topologically protected ONR.

%
The mechanism can also be generalized to a nonreciprocal amplifier \cite{Bahari2017}, where the gain medium can be incorporated efficiently in compact 1D materials. Furthermore, our scheme based on 1D SSH model can be directly generalized to 2D SSH models \cite{Liu2017} and the associated higher-order topological matters \cite{Benalcazar2017,Ezawa2018}, which exhibit different Zak phases along different directions. In particular, we can design the ONR between any pair of ports within a multi-port 2D device.

\section*{Code availability}
The simulation code is available from the corresponding authors upon reasonable request.

%

\section*{Competing interests}
The authors declare no competing interests.

\section*{Methods}

\subsection*{Experimental setup}

The experiment is carried out in a 2-cm long vapor cell that contains natural abundance rubidium atoms. A titanium sapphire laser (Spectra-Physics, Matisse CX, 1 mm $e^{-2}$ beam width) generates two counter-propagating coupling fields. The $\pm x$-directional probe field, generated by a diode laser (Toptica, DLC TA PRO 795,  0.5 mm $e^{-2}$ beam width), intersects with the coupling fields at a small angle ($\sim 0.3^\circ$) in the vapor cell. Transmission spectra are recorded by two photodiode detectors in different directions. The experimental data in Fig.~\ref{fig2} and Extended Data Fig.~\ref{Extdata1} are averaged over the inteval $[-\Gamma/2,\Gamma/2]$ relative to $\delta=0$. Here $\Gamma=2\pi\times 5.7$ MHz is the decay rate of the state $|b\rangle$.

\subsection*{The SSH SLs}

We notice that the Hamiltonian $H_{c}(x)$ resembles a band structure in a real-space Brillouin zone. By Fourier transforming $H_c$ to momentum space, we obtain an SSH Hamiltonian of the SLs \cite{Wang2015,Chen2018,Mi2021},  
\begin{equation}
    H_{\text{SL}} =
            \sum_n \left[\Omega_1 b^{\dagger}_{2n+1} a_{2n}
        + \Omega_2 b^{\dagger}_{2n-1} a_{2n}
        + \text{H.c.}\right],
\end{equation}
where $a^{\dagger}_{2n}=1/\sqrt{N_0}\sum_l e^{2ink x_l}|a_l\rangle\langle g_l|$  and $b^\dagger_{2n+1}=1/\sqrt{N_0}\sum_l e^{i(2n+1)k x_l}|b_l\rangle\langle g_l|$ are creation operators of the timed Dicke states containing $2n$ and $2n+1$ quanta of momentum $\hbar k$, $N_0$ is the total number of atoms, and $x_l$ is the position of the $l$th atom with atomic levels $|a_l\rangle$, $|b_l\rangle$, and $|g_l\rangle$. The intra-(inter-)cell hopping between the $a_{2n}$ and $b_{2n+1}$ ($b_{2n-1}$) sites describes the coupling between the atomic ensemble and the plane wave coupling field propagating in the $+x$ ($-x$) direction with a Rabi frequency $\Omega_1$ ($\Omega_2$).
The atomic motion introduces a momentum-space linear potential, 
\begin{equation}
    V=\sum_n kv \left[(2n+1)b^\dagger_{2n+1}b_{2n+1}+2na^\dagger_{2n}a_{2n}\right],
\end{equation} 
owing to the frequency of the $\pm x$ coupling field being shifted by $\mp kv$ in the reference frame of the atoms. 

For a small velocity such that $|kv|\ll\bar{\Omega}$, which is the average Rabi frequency $\bar{\Omega}=2/\lambda\int_{\lambda/2} dx|\Omega(x)|$, $V$ acts as a perturbation to $H_\text{SL}$. In this limit, the energy spectra are WSLs \cite{Zak1989,Xiao2010,Mao2022,Sundaram1999,King1993,Resta1994}, $\varepsilon\approx \pm \bar{\Omega} + (n+\theta/\pi)kv$ with $n$ being an integer. The WSL spectra are linear to the Zak phase $\partial \varepsilon/\partial (kv)\propto n+\theta/\pi$, and the corresponding eigenstates are localized in the $n$th momentum-space unit cell. In the large velocity force limit $|kv|\gg \bar{\Omega}$, we can ignore $H_{\text{SL}}$ and the eigenspectra are dominated by the linear potential, $\partial\varepsilon/\partial (kv)\approx n$, where the Zak phase contribution disappears.

\subsection*{Temperature dependence}

The transmission of the probe field is given by $T=\exp(-AL)$, where $L$ is the length of the rubidium vapor cell and the absorption coefficient \cite{Paladugu2024} is defined as
\begin{eqnarray}
        A(\delta) = \frac{N\mu^2\pi}{2\epsilon_0}\int dvP(v)\sum_l\delta_D(\delta+\varepsilon_l)|\psi_l^1|^2.
\end{eqnarray}
Here $\mu$ is the dipole moment of the transition, $N$ is the atomic density, $\epsilon_0$ is the permittivity of vacuum, $\int dv P(v) = 1$ represents the normalized Maxwell distribution, $\delta_D(\cdot)$ is the Dirac delta function, and $|\psi_l^1|^2=\langle \psi_l|b^{\dagger}_1b_1|\psi_l\rangle$ describes the probability at site $b_1$ for the eigenstate $|\psi_l\rangle$, which satisfies the Schr\"{o}dinger equation $(H_{\text{SL}}+V)|\psi_l\rangle = \varepsilon_l|\psi_l\rangle$. The absorption coefficients at  zero detuning are approximately,
\begin{eqnarray}
        A(\delta) \approx
        \begin{cases}
            0, \quad $\text{when}$ \quad \Delta=\pm0.5kv,\\  
            \frac{N\mu^2\pi\Gamma}{2\epsilon_0 k}
            \left[P(\frac{\bar\Omega}{\sqrt{2}k})+P(-\frac{\bar\Omega}{\sqrt{2}k})\right], \quad $\text{when}$ \quad  \Delta=\pm1.5kv. 
        \end{cases}
\end{eqnarray}


The temperature of the vapor cell has two effects on the absorption coefficients, the width of the Maxwell distribution $P(v)$ and the atomic density $N$. In our experiment, we change the temperature over a range of tens of degrees centigrade, during which the width of the velocity distribution can be well approximated as a constant (Extended Data Fig.~\ref{Extdata1}(a)). Therefore, we can rewrite the transmission as $T=e^{-\alpha N(T_K)}$, where $T_K$ is the temperature and $\alpha>0$ is a constant independent of temperature. Finally, we conclude that the range of the transmission contrast scales as $[1-e^{-\alpha N(T_K)}]/[1+e^{-\alpha N(T_K)}]$, which approaches to 1 with increasing temperature, as shown in Extended Data Fig.~\ref{Extdata1}(b).

\renewcommand{\figurename}{Extended Data Fig.}

\setcounter{figure}{0}

\newpage

\begin{figure*}[htbp]
    \centering
    \includegraphics[width=1\textwidth]{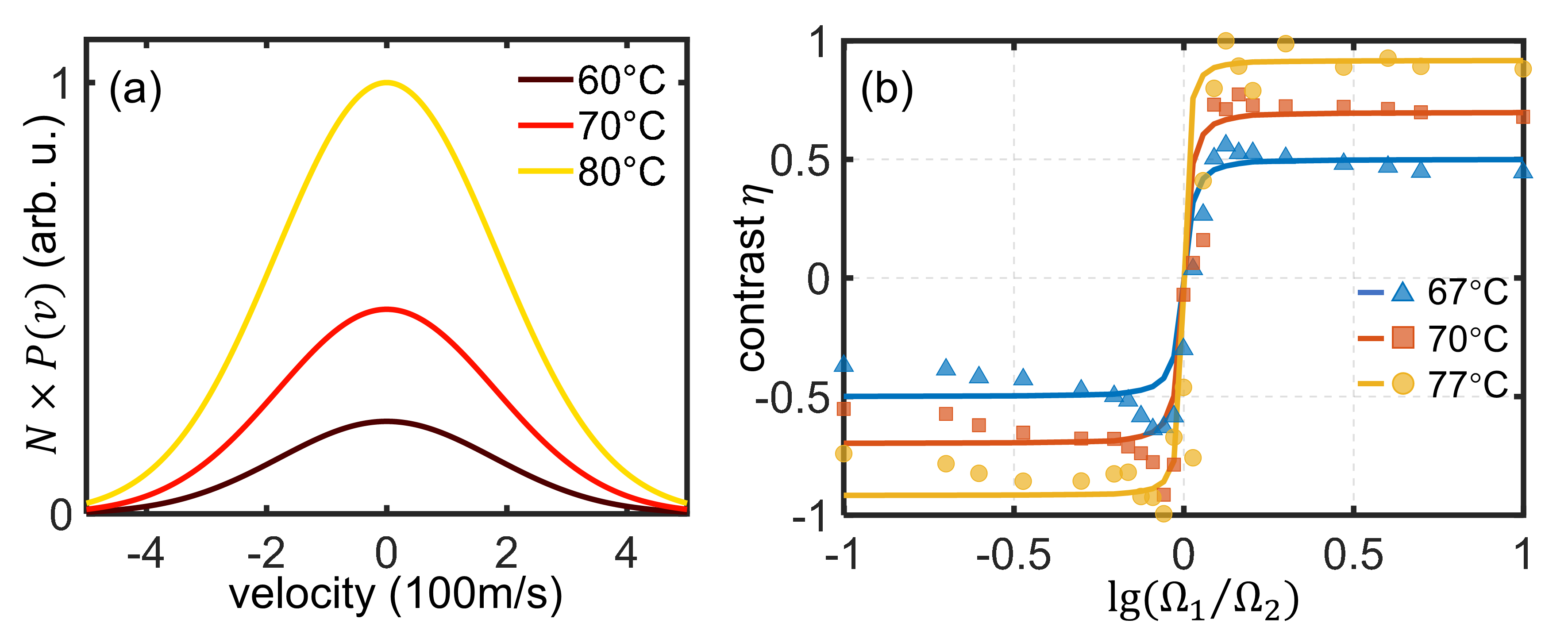}
    \caption{
        (a) The atomic velocity distribution.
        (b) The transmission contrast at different temperatures. The lines are numerical results and the dots are experimental data. For each data point, the larger Rabi frequency is set as 136 MHz. At the topological transition point, we set $\Omega_1=\Omega_2=136$ MHz.}
        \label{Extdata1}
\end{figure*}

\end{document}